# Effects of Reynolds Number and Stokes Number on Particle-pair Relative Velocity in Isotropic Turbulence: A Systematic Experimental Study

Zhongwang Dou[a*], Andrew D. Bragg[b], Adam L. Hammond[a], Zach Liang[a], Lance R. Collins[c], Hui Meng[a**]

[a] Department of Mechanical and Aerospace Engineering, University at Buffalo, Buffalo, NY, 14260, United States

[b] Department of Civil and Environmental Engineering, Duke University, Durham, NC, 27708, United States

[c] Sibley School of Mechanical and Aerospace Engineering, Cornell University, Ithaca, NY, 14850, United States

* Current Address: 319 Latrobe Hall, Johns Hopkins University, Baltimore, MD, 21218

** Corresponding Author:

Hui Meng, Ph.D.

Department of Mechanical and Aerospace Engineering

University at Buffalo

Buffalo, NY 14260, United States

Telephone: (716) 645-1458

FAX: (716) 645-2883

Email: huimeng@buffalo.edu

**Abstract**

The effects of Reynolds number ($R_\lambda$) and Stokes number ($St$) on particle-pair relative velocity (RV) were investigated systematically using a recently developed planar four-frame particle tracking technique in a novel homogeneous and isotropic turbulence chamber. We compare measurement results with DNS, verifying if conclusions in DNS at simplified conditions and limited $R_\lambda$ still valid in reality. Two experiments were performed: varying $R_\lambda$ between 246 and 357 at six $St$ values, and varying $St$ between 0.02 and 4.63 at five $R_\lambda$ values. Measured mean inward particle-pair RV $\langle w_r^- \rangle$ as a function of separation distance $r$ were compared against DNS under closely matched conditions. At all experimental conditions, an excellent agreement was achieved except when particle separation distance $r \lesssim 10\eta$ ($\eta$: Kolmogorov length scale), where experimental $\langle w_r^- \rangle$ was consistently higher, possibly due to particle polydispersity and finite laser thickness in experiment (Dou et al., 2017). At any fixed $St$, $\langle w_r^- \rangle$ was essentially independent of $R_\lambda$, echoing DNS finding by Ireland et al. (2016a). At any fixed $R_\lambda$, $\langle w_r^- \rangle$ increased with $St$ at small $r$, showing dominance of path-history effect in the dissipation range when $St \geq O(1)$, but decreased with $St$ at large $r$, indicating dominance of inertial filtering. We further compared $\langle w_r^- \rangle$ and RV variance $\langle w_r^2 \rangle$ from experiments against DNS and theoretical predictions by Pan and Padoan (2010). For $St \lesssim 1$, experimental $\langle w_r^- \rangle$ and $\langle w_r^2 \rangle$ matched them well at $r \gtrsim 10\eta$, but were higher than both DNS and theory at $r \lesssim 10\eta$. For $St \gtrsim 1$, $\langle w_r^- \rangle$ from all three matched well except for $r \lesssim 10\eta$, in which experimental values were higher, while $\langle w_r^2 \rangle$ from experiment and DNS were much higher than theoretical predictions. We discuss potential causes of these discrepancies. What this study shows is the first experimental validation of $R_\lambda$ and $St$ effect on inertial particle-pair $\langle w_r^- \rangle$ in homogeneous and isotropic turbulence.

## 1. Introduction

The study of turbulence-enhanced inertial particle collision in isotropic turbulence could improve our understanding and modeling of many particle-laden turbulent flows in nature. For example, water droplet development in the "size-gap" in warm clouds is believed to be dominated by turbulence-induced particle collision (Shaw, 2003). The formation of planetesimals in protoplanetary disks is believed to begin with small dust grains that collide and coalesce in turbulent protoplanetary nebulae (Dullemond and Dominik, 2005; Zsom et al., 2010).

These phenomena are associated with high Taylor-microscale Reynolds numbers ($R_\lambda$) and a wide range of particle Stokes numbers ($St$). For example, in cumulus clouds $R_\lambda = O(10^5)$, and $St \approx 0.01 - 2$ (Ayala et al., 2008; Siebert et al., 2006), while in protoplanetary nebulae $R_\lambda = O(10^4) - O(10^6)$ and $St \approx O(1) - O(10^3)$ (Cuzzi et al., 2001). However, since high $R_\lambda$ can neither be generated through typical laboratory facilities nor through high performance computing, many researchers have studied inertial particle collisions in isotropic turbulence at relatively low Reynolds numbers ($R_\lambda = 50 - 500$). In order to understand if findings from lower Reynolds numbers are applicable to naturally occurring high Reynolds number turbulent flows, it is necessary to quantify the effect of $R_\lambda$ on particle collision rate in isotropic turbulence.

On the other hand, particle collision is also influenced by particle inertia in relation to turbulent flow, or the Stokes number $St$ (Ireland et al., 2016a; Pan and Padoan, 2013; Sundaram and Collins, 1997). In different $St$ ranges, the effect of $St$ on particle collision may be dominated by different mechanisms (Bragg and Collins, 2014a, b; Bragg et al., 2015b). It is necessary to quantify, for a wide $St$ range, the effect of $St$ on particle collision in isotropic turbulence.

While particle collision itself is difficult to model numerically and capture experimentally, the collision rate of inertial particles is known to depend upon two distinct phenomena, namely particle clustering and particle-pair relative motion in isotropic turbulence. In the dilute limit, the collision kernel of monodispersed inertial particles can be expressed as (Sundaram and Collins, 1997; Wang et al., 2000)

$$K(d) = 4\pi d^2 g(d)\langle w_r(d)^-\rangle, \qquad (1)$$

where $K(d)$ is the collision kernel, $d$ is the diameter of the particle, $g(r)$ is the radial distribution function (RDF) of particles, and $\langle w_r(r)^-\rangle$ is the particle-pair mean inward radial relative velocity (RV). In Eq. (1), the separation distance $r$ between two particles is taken at contact, i.e. $r = d$. The influence of $R_\lambda$ and $St$ on particle collision rates can be investigated by considering how $R_\lambda$ and $St$ affect the RDF and particle-pair RV.

Many studies have examined the RDF in isotropic turbulence (Bragg and Collins, 2014a; Bragg et al., 2015a; Collins and Keswani, 2004; Eaton and Fessler, 1994; Salazar et al., 2008; Wu et al., 2017), and more recently, particle-pair RV (Bec et al., 2010; Gustavsson and Mehlig, 2011; Ireland et al., 2016a; Salazar and Collins, 2012b). From these theoretical and numerical studies, a consensus has emerged concerning the effect of $St$ on the RV, presented in Bragg and Collins (2014b) and Ireland et al. (2016a), which we shall now summarize for the purposes of making the present paper self-contained (we refer the reader to the cited papers for more detailed and precise explanations).

Turbulence affects the particle relative motion through three mechanisms: preferential sampling, path-history, and inertia filtering effects. Here, we define $\tau_r$ as the turbulence eddy turnover timescale at the particle-pair separation $r$, and $\tau_p$ as the particle response time. First, when $\tau_p \ll \tau_r$, the particle motion is only slightly perturbed relative to that of fluid particles. In this regime, the inertial particle-pair RV differs from the fluid particle RV only because of the *preferential sampling mechanism*, which describes the tendency of inertial particles to preferentially sample certain regions of the fluid velocity field, unlike fluid particles that uniformly (ergodically) sample the underlying velocity field. Except at low Reynolds numbers, the preferential sampling effect causes inertial particle-pair RV to be reduced compared to those of fluid particles (Ireland et al., 2016b).

Second, when $\tau_p = O(\tau_r)$, the inertial particle-pairs retain a finite memory of the fluid velocity differences they have experienced along their path-history. This gives rise to the *path-history effect* on particle-pair RV. At sub-integral scales, the fluid velocity differences increase with increasing separation, on average (though in the dissipation regime this is true instantaneously). As a result, on average there is an asymmetry in the nature of the turbulence experienced by particle-pairs that approach compared with those that separate, and this asymmetry can strongly affect the inertial particle-pair RV statistics. In particular, it leads to an increase in their relative velocities compared to those of fluid particles. It should be noted, however, that the importance of this path-history mechanism depends not only upon the ratio $\tau_p/\tau_r$, but also upon $r$. The effect is most profound in the dissipation range, and its effect decreases with increasing $r$. This is because the fluid velocity differences increase with $r$ more weakly as $r$ is increased, and indeed become independent of $r$ for $r \gg L$, where the path-history mechanism vanishes at these scales. This path-history mechanism is the same mechanism that leads to "caustics", "the sling effect", and "random, uncorrelated motion" (Bragg and Collins, 2014b; Falkovich and Pumir, 2007; Ijzermans et al., 2010; Wilkinson et al., 2006).

Finally, when $\tau_p = O(\tau_r)$, the particle inertia also gives rise to the *filtering mechanism* (Ayyalasomayajula et al., 2008; Bec et al., 2006; Ireland et al., 2016a; Ireland et al., 2016b), which describes the modulated response of the particles to fluctuations in the turbulent velocity field because of their inertia. This mechanism always causes the inertial particle-pair RV to decrease relative to that of fluid particles. The filtering mechanism operates at all $r$ when $\tau_p \approx \tau_r$. However, its importance compared with the path-history effect depends upon $r$. For $r \lesssim \eta$, the path-history effect dominates the inertial particle-pair RV, while for $r \gg L$, the filtering effect completely dominates the inertial particle RV, since at large scales the path-history mechanism vanishes.

The ways in which Stokes number and Reynolds number affect inertial particle-pair RV are through their influence on these three mechanisms. The Stokes number, being a measure of particle response time $\tau_p$ (normalized by Kolmogorov time scale), affects the relative importance of these different mechanisms. At small $St$ ($St \ll 1$), the preferential sampling effect dominate particle-pair RV, leading to slightly lower RV values compare to fluid particle (Salazar and Collins, 2012a); At large $St$ ($St \approx O(1)$), the path-history dominate small $r$, while the inertia filtering dominate large $r$, resulting an increase and decrease of particle-pair RV, respectively.

We note that when the path-history effect is weak, the inertial particle relative velocities may be estimated in terms of the correlation function of the particle and fluid relative velocities at sepration $r$. This reflects the fact that in this case, the RVs of the particles are determined by how well correlated the particle motion is with the local fluid velocity field and its associated structure. The following models attempt to predict the particle-pair RV based on this idea: Laviéville et al. (1995), Laviéville et al. (1997), Simonin (2000), and Zaichik and Alipchenkov (2009).

Compared with the effect of $St$, the effect of Reynolds number on particle-pair RV is far less understood. There are essentially two kinds of Reynolds number effects. One is the "trivial" effect which describes the fact that as the Reynolds number is increased, the scale separation (spatial and temporal) of the turbulence increases. Theoretical models of inertial particle-pair RV captured this trivial effect (Pan and Padoan, 2010; Zaichik and Alipchenkov, 2009). However, the more complex question concerns how the RV statistics might be affected by the "non-trivial" Reynolds number effects, e.g. through internal intermittency and potential modifications to the spatial-temporal structure of the flow. Theoretical studies by Falkovich et al. (2002) conjectured that an increase of the Reynolds number could lead to larger particle-pair RV through the enhanced intermittency of the fluid velocity field. Although this is undoubtedly true for higher-order statistical quantities (associated with "extreme events" in the turbulence), for the lower-order quantities relevant to particle collisions in turbulence, recent numerical studies from Bragg et al. (2016a), Bec et al. (2010), and Rosa et al. (2013) have found that the effect of Reynolds number is very week.

However, those numerical simulations or theoretical interpretations were obtained under simplified and limited condition, i.e. particle size are monodispersed, particle-turbulence interaction are one-way coupling, the Basset history forces, nonlinear drag, and hydrodynamic interactions are ignored. In the natural phenomena or real engineering applications, those terms may not be avoidable. In order to know if findings in simulations were applicable under more complex circumstances, experimental investigation of these finds are clearly needed.

These findings from theoretical and numerical studies have not yet been verified experimentally, except for one case – particle-pair RV at $R_\lambda \approx 180$ in a small range of low $St$ $(0.05 - 0.5)$ (Saw et al., 2014). Using high-speed particle tracking in an enclosed turbulence chamber, Saw et al. (2014) experimentally found that $\langle w_r^- \rangle$ increases with $St$ when $r$ is comparable to $\eta$, the Kolmogorov length scale, which agreed with their DNS results. However, systematic experimental explorations of $R_\lambda$ and $St$ effects on particle-pair RV are still lacking, and it is important to comprehensively verify these findings from theory and simulation through experiment.

The limited availability of high-speed particle tracking instrument and the challenging nature of these experiments account for the dearth of experimental measurement of particle-pair RV. In our preliminary study, de Jong et al. (2010) measured particle-pair RV using double-exposure holographic particle imaging in a cubic turbulence chamber. Their measurement exhibited large errors at large RV values due to particle positioning uncertainty along the depth direction caused by the limited angular aperture of digital holography, as well as due to particle pairing error based on only two exposures to track particles (Dou et al. (2017). More recently, Dou et al. (2017) demonstrated greatly improved particle-pair RV measurement using a novel planar four-frame particle tracking velocimetry (planar 4F-PTV) technique in a new fan-driven "soccer ball" shaped homogeneous and isotropic turbulence (HIT) chamber. They measured particle-pair RV at two $St$ values ($St$=0.09 and 3.46) and a fixed $R_\lambda = 357$ and reported an excellent match between experimental and DNS results from $10 < r/\eta < 60$. Both experiment and DNS show that particle-pair RV increases with $St$ in the dissipation range and decrease with $St$ in the inertial range. However, when $r/\eta \lesssim 10$, the experimentally measured $\langle w_r^- \rangle$ was higher than DNS. They attributed this difference to the measurement uncertainty caused by polydispersion of particles and finite light sheet thickness in the experiment, with DNS being run for 3D and monodispersed particles. These authors only examined two $St$ conditions at a single $R_\lambda$ and did not address the effect of changing $R_\lambda$ on particle-pair RV.

In this paper, we report a systematic experimental investigation of the effects of Reynolds number and Stokes number on mean inward particle-pair RV, based upon the pilot study by Dou et al. (2017). Using the 4F-PTV technique and the HIT chamber descried by Dou et al. (2017), we systematically but independently varied $R_\lambda$ and $St$ over wide ranges and measured $\langle w_r^- \rangle$ under a large number of experimental conditions. Furthermore, in order to examine if findings from numerical simulations and theoretical interpretations still stands at more complex circumstances, we compared experimental results of variance of RV $\langle w_r^2 \rangle$ and mean inward RV $\langle w_r^- \rangle$ against DNS by Ireland et al. (2016a) and theoretical prediction by Pan and Padoan (2010).

## 2. Experimental Method

### *2.1. Flow Facility and Measurement Technique*

Our experiments were conducted in an enclosed, one-meter-diameter, truncated icosahedron chamber, which generates high-Reynolds-number turbulence homogeneous and isotropic turbulence in the center region through 20 symmetrically distributed actuators along 10 axes (Dou, 2017; Dou et al., 2016). This flow chamber produces homogeneous and isotropic turbulence (with near zero-mean flow and a maximum $R_\lambda$ of 384) in a spherical volume in the center of at least 48 mm in diameter with minimal gravitational effects on the particles. A 0.2" diameter hole was placed at the bottom of the facility to allow particle injection through an attached injector and compressed air was employed for particle injection before each test run. A thermocouple probe (Type K) was inserted into the chamber near the back of a fan to monitor temperature fluctuations. We further modified the chamber to remove static electric charge buildup on particles by coating the interior surfaces of chamber with carbon conductive paint (Dou et al., 2017).

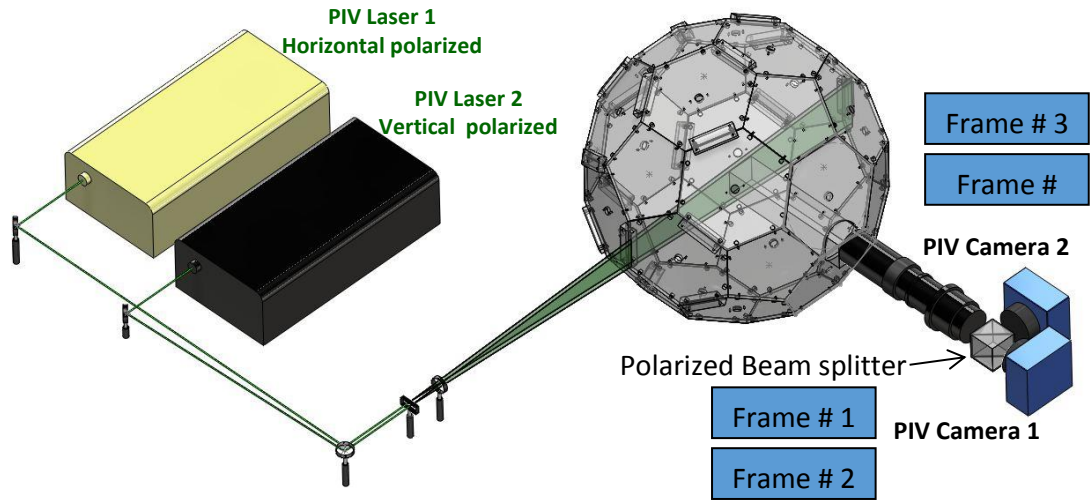

Figure 1: The setup of measuring inertial particle-pair relative velocity using planar four frame particle tracking velocimetry in a homogeneous and isotropic turbulence chamber.

We use in-house developed planar 4F-PTV technique for the measurement of particle-pair RV in this study, shown in figure 1. The detail of 4F-PTV has been described by Dou et al. (2017). Briefly, the 4F-PTV system employs two double-exposure PIV cameras and two double-pulse PIV lasers. PIV systems combined together spatially and temporally. Spatially, two PIV lasers beam are overlapped to illuminate the same flow region, and the two PIV cameras were orthogonally placed next to a polarized beam splitter to focus on the same flow region. Temporally, a timing unit synthetizes the two PIV lasers and cameras. Two PIV lasers, each generates two short laser pluses, consecutively generated 4 laser pluses with the same time intervals. The first PIV camera captured the first and second particle images, while the second PIV camera captured the third and fourth particle images. Since polarized beam splitters were used and laser pluses from the two PIV lasers were horizontally and vertically polarized, these two double-exposures were not contaminated with each other. This high-speed planar 4F-PTV technique has the advantage of a very short time interval (on the order of microseconds) and accurate particle pairing capability that enables high-speed particle tracking over four consecutive frames using routine lab equipment (i.e. regular PIV systems).

*2.2 Experimental Conditions*

The turbulence Taylor microscale Reynolds number is expressed as

$$R_\lambda = u'^2 \sqrt{15/\upsilon\varepsilon}, \tag{2}$$

where $u'$ is the turbulence strength, $\upsilon$ is the kinematic viscosity and $\varepsilon$ is the turbulence kinetic energy dissipation rate. The inertia of particles in isotropic turbulence is quantified by Particle Stokes number, $St \equiv \tau_p/\tau_\eta$, the ratio of particle response time, $\tau_p = \rho_p d^2/(18\upsilon\rho_f)$, to turbulence Kolmogorov time scale, $\tau_\eta = \sqrt{\upsilon/\varepsilon}$. It can be shown that

$$St = \frac{\rho_p d^2 \sqrt{\varepsilon}}{18\upsilon^{3/2} \rho_f}. \tag{3}$$

Here $\rho_p$ and $\rho_f$ are the particle and fluid densities, respectively, and $d$ is particle diameter.

In order to generate a wide range of $St$ at different $R_\lambda$ in our HIT chamber, we used two types of commercial particles: the low-density glass bubbles (3M Inc., K25, $\rho_p = 0.25 g/cc$) and the high-density glass bubbles (3M Inc., S60, $\rho_p = 0.60 g/cc$). Both low- and high-density glass bubbles were run through a series of sieves (ASTM E161 compliant) to generate diameter ranges ($5-10\ \mu m$, $15-20\ \mu m$, $25-32\ \mu m$, $32-38\ \mu m$, and $38-45\ \mu m$) in a particle separation instrument (GilSonic UltraSiever GA-8). Based on the available flow conditions in the fan-driven HIT chamber ($R_\lambda = 246, 277, 324, 334,$ and $357$), as well as the obtainable particle density and diameter, we list all possible experimental conditions in table 1. Note that the particle samples in the experiments are narrowly polydispersed, and the effective $St$ of a group of polydispersed particles can be slightly different than their mean $St$ (Zaichik et al., 2006).

TABLE 1. Experimental conditions for particle-laden turbulence in the HIT chamber used in this study. In some particle size ranges, two particle densities were required. In total, 40 unique experimental conditions were used.

| Flow Condition | $R_\lambda$ | | 246 | 277 | 324 | 334 | 357 |
|---|---|---|---|---|---|---|---|
| | $\varepsilon (m^2/s^3)$ | | 3.6 | 9.2 | 16.5 | 27.04 | 35.89 |
| Particle Characteristics | Dia. ($\mu m$) | Density ($g/cc$) | Average $St$ | | | | |
| | 5-10 | 0.25 | 0.02 | 0.04 | 0.05 | 0.06 | 0.07 |
| | 15-20 | 0.25 | 0.11 | 0.18 | 0.24 | 0.30 | 0.35 |
| | 20-25 | 0.25 | 0.18 | 0.29 | 0.39 | 0.49 | 0.57 |
| | 25-32 | 0.60 | 0.70 | 1.11 | 1.49 | 1.91 | 2.20 |
| | | 0.25 | 0.29 | 0.46 | 0.62 | 0.80 | 0.92 |
| | 32-38 | 0.60 | 1.04 | 1.67 | 2.23 | 2.86 | 3.29 |
| | | 0.25 | 0.43 | 0.69 | 0.93 | 1.19 | 1.37 |
| | 38-45 | 0.60 | 1.47 | 2.34 | 3.14 | 4.02 | 4.63 |

*2.3 Experimental Design*

In order to independently study the effects of $R_\lambda$ and $St$ on particle-pair RV statistics, we conducted two series of experiments: Experiment A (sweeping $R_\lambda$ at fixed $St$) and Experiment B (sweeping $St$ at fixed $R_\lambda$). The two variables were controlled independently by changing the flow conditions and particle characteristics. To compare with DNS results of particle-pair RV by Ireland et al. (2016a), ideally we should match the $R_\lambda$ and $St$ of our experiments with their DNS conditions. In practice, the available $R_\lambda$ values for experiments are determined by the flow facility (table 1), and the whole range of experimental $R_\lambda$ (247~357) falls between two of the $R_\lambda$ values in DNS by Ireland et al. (2016a), namely 224 and 398. Since there was essentially no difference in mean inward RV, $\langle w_r^- \rangle$ and variance of RV, $\langle w_r^2 \rangle$ under these two $R_\lambda$ over the range (small and intermedia $St$ and $r$) of interests in the DNS results (see figure 15 of Ireland et al. (2016a), we used their DNS results under $R_\lambda = 398$ and $St = 0, 0.05, 0.3, 0.5, 1, 2, 3$ and 10 to compare with our experimental measurements.

**TABLE 2.** Experiment A (Sweeping $R_\lambda$): list of the nominal $St$, actual $St$, and swept $R_\lambda$ conditions for six test groups (A1 – A6).

| Test Group | Nominal $St$ (to match DNS) | Actual $St$ mean, range | $R_\lambda$ |
|---|---|---|---|
| A1 | ~0.05 | 0.05, [0.02 - 0.07] | 247, 277, 324, 334, 357 |
| A2 | ~0.3 | 0.29, [0.24 - 0.35] | 247, 277, 324, 334, 357 |
| A3 | ~0.5 | 0.51, [0.43 - 0.62] | 247, 277, 324, 334, 357 |
| A4 | ~1.0 | 1.04, [0.92 - 1.19] | 247, 277, 324, 334, 357 |
| A5 | ~2.0 | 2.17, [1.91 - 2.34] | 277, 324, 334, 357 |
| A6 | ~3.0 | 3.10, [2.86 - 3.29] | 324, 334, 357 |

Note that for $St$~2.0, the lowest $R_\lambda$ can be obtained is 277, and for $St$~3.0, the lowest $R_\lambda$ can be obtained is 324.

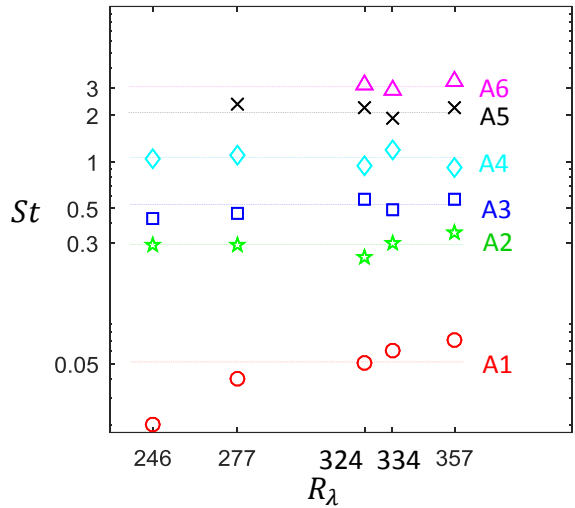

**FIGURE 2.** Graphical representation of table 2: Experimental conditions for Test Groups A1 – A6. Each color represents one sweeping $R_\lambda$ test group at a certain $St$, while each point represents one unique experimental condition.

***Experiment A: Sweeping $R_\lambda$ at fixed $St$:*** To study the effect of $R_\lambda$ on particle-pair RV, we experimentally swept $R_\lambda$ at six target $St$ values: $St = 0.05, 0.3, 0.5, 1, 2,$ and 3. In practice, it is tricky to sweep $R_\lambda$ while keeping $St$ constant, since an increase of $R_\lambda$ is accompanied by a decrease of $\tau_\eta$ and thereby an increase of $St$ for the same particles. To maintain a nominal $St$ value, increasing $R_\lambda$ requires decreasing particle time scale $\tau_p$ accordingly. This requires either the particle diameter ($d$) or density ($\rho$) or both to be reduced accordingly as well. Since the available particle densities and diameter ranges are limited, we had to choose different particles for different $R_\lambda$ in order to get $St$ as close to the nominal $St$ values as possible.

For this purpose, a part of experimental conditions in table 1 are reorganized by the nominal $St$ values, with $R_\lambda$ being the variable, and denoted as A1, A2, A3, A4, A5, and A6 as shown in table 2. Each test group held constant nominal $St$ while changing $R_\lambda$ from 246 to 357. figure 2 graphically illustrates the conditions of these test groups, showing that fluctuations of the actual $St$ were within 20% of the nominal $St$ values except for Test Group A1.

**TABLE 3.** Experiment B (Sweeping $St$): list of the fixed $R_\lambda$ and swept $St$ conditions for five test groups (B1 – B6).

| Test Group | $R_\lambda$ | $St$ | | | | | | | |
|---|---|---|---|---|---|---|---|---|---|
| B1 | 246 | 0.02 | 0.11 | 0.18 | 0.29 | 0.43 | 0.70 | 1.04 | 1.47 |
| B2 | 277 | 0.04 | 0.18 | 0.29 | 0.46 | 0.69 | 1.11 | 1.67 | 2.34 |
| B3 | 324 | 0.05 | 0.24 | 0.39 | 0.62 | 0.93 | 1.49 | 2.23 | 3.14 |
| B4 | 334 | 0.06 | 0.30 | 0.49 | 0.80 | 1.19 | 1.91 | 2.86 | 4.02 |
| B5 | 357 | 0.07 | 0.35 | 0.57 | 0.92 | 1.37 | 2.20 | 3.29 | 4.63 |

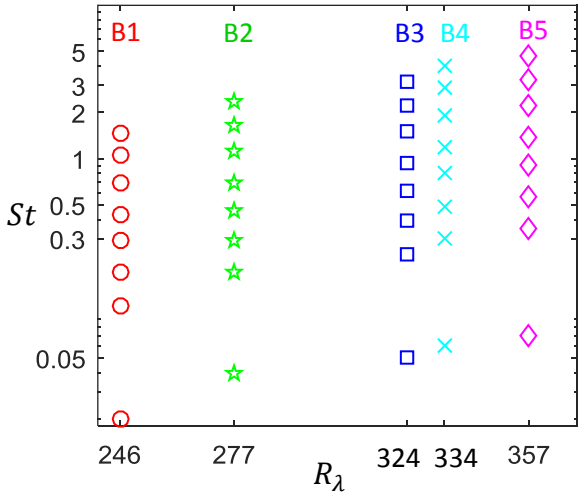

**FIGURE 3**. Graphical representation of Table 3: Experimental conditions for Tests Groups B1 – B5. Each color represents one sweeping $St$ test group at a certain $R_\lambda$, while each point represents one unique experimental condition.

***Experiment B: Sweeping $St$ at fixed $R_\lambda$***: To study the effect of $St$ on particle-pair RV, we experimentally swept $St$ at five $R_\lambda$ values: $R_\lambda$ = 246, 377, 324, 334, and 357. To change $St$ at fixed $R_\lambda$, the flow condition was held constant and different particles were used. For each flow condition and thus $R_\lambda$, all particle size and density permutations listed in table 1 were used to acquire eight $St$ values. The test groups for five different $R_\lambda$ are denoted as B1, B2, B3, B4, and B5, and shown in table 3 and figure 3.

## 2.4 Experimental Procedure

For all of the 40 experiment conditions (table 1), we performed particle-pair RV measurement using the planar 4F-PTV technique in the HIT chamber. Statistics of the particle motion were only collected and computed once the turbulence had reached a statistically stationary state, plus an additional ten large eddy time scales. Furthermore, the particle number density in the turbulence chamber is estimated to be below 50 per cube centimeter (particle volume fraction of order $10^{-6}$) such that the system can be considered dilute. Under each test condition, we obtained 10, 000 quadruple-exposure particle images in 20 runs of 500 consecutive quadruple pulses, repeated at 5 - 9.5 Hz. The time interval in the quadruple-exposure varied between 38 $\mu s$ and 91 $\mu s$ for different $R_\lambda$. In each quadruple-exposure, we obtained individual particle positions and velocities. We calculated particle-pair RV as

$$w_r(r) = (\boldsymbol{v_A} - \boldsymbol{v_B}) \cdot \frac{\boldsymbol{r}}{|\boldsymbol{r}|}, \tag{4}$$

where $\boldsymbol{v_A}$ and $\boldsymbol{v_B}$ are the velocity vectors of Particle A and B respectively, and $\boldsymbol{r}$ is the distance vector from Particle A to Particle B. The calculated particle-pair RV, $w_r(r)$, were binned with the particle separation distance, $r$, at increments of $\eta$. Particle-pair RV statistic of mean inward particle-pair RV, $\langle w_r^- \rangle$, and variance of particle-pair RV, $\langle w_r^2 \rangle$, were obtained over a wide range of $r$. We calculated all experimental uncertainty following the procedure reported in Appendix B of Dou et al. (2017). In the final calculation, we normalized $r$, $\langle w_r^- \rangle$, and $\langle w_r^2 \rangle$ by Kolmogorov length scale $\eta$, Kolmogorov velocity scale $u_\eta$, and the square of Kolmogorov velocity scale $u_\eta^2$, respectively. We then plotted these normalized quantities based on the test groups in Experiments A and B.

*2.5 Monte Carlo Analysis to Account for Out-of-Plane Components of Particle-Pair RV*

The planar 4F-PTV technique measures the particle-pair RV within a laser light sheet that has a finite thickness of $8\eta$. The laser thickness is required to contain a sufficient number of particles within the light sheet over the four consecutive exposures for accurate particle tracking. However, the planar PTV measures only in-plane particle-pair separation and plane relative velocity, omitting the out-of-plane components, which brings about uncertainty of RV measurement when the particle separation distance is small. Dou et al. (2017) evaluated the finite laser thickness effect and shown that it causes an overestimation of particle-pair RV in small $r$. When the in-plane particle separation distance $r$ is larger than the laser thickness, the difference in $\langle w_r^- \rangle$ dropped down to below 10%. They further attempted to correct the omission of the out-of-plane components of RV using Monte Carlo analysis (MCA) on their experimental data, assuming the same particle distribution in the out-of-plane direction as in the transverse directions based on isotropy. This correction procedure was able to eliminate the finite laser thickness effect on RV measurement for $r \gtrsim 5\eta$.

We have applied the MCA correction all the experimental results of $\langle w_r^- \rangle$ in both Experiments A and B to account for the finite laser thickness, and compared the results after and before the out-of-plane correction.

## 3. Results and Discussions

We present all the raw experimental data from Experiments A and B in this section. After the MCA correction for the out-of-plane components, all the plots of $\langle w_r^- \rangle$ against $r$ remained the same for $r \geq 15\eta$ (indicating negligible effect), moved down by 5-10% for $r = 5 - 15\eta$ (successful correction), and 3-5% for $r \leq 5\eta$ (limited correction), which are consistent with report by Dou et al. (2017). Because the MCA correction method itself contains assumptions and simulation, and because the difference after correction is minimal and not enough to change the trends of $\langle w_r^- \rangle$ versus $r$, $\langle w_r^- \rangle$ versus $R_\lambda$, and $\langle w_r^- \rangle$ versus $St$, we choose to present the raw experimental data without the MCA correction.

### 3.1 Mean Inward Particle-pair RV

In figure 4 and 5, we plot the mean inward particle-pair RV, $\langle w_r^- \rangle$, versus particle-pair separation distances $r$ for some of the experimental conditions to give an overview of the general trend using both linear and log scales. The rows represent the sweeping of $R_\lambda$ in Experiment A, while the columns represent the sweeping of $St$ in Experiment B. Thirteen unique experimental conditions of Experiment B (sweeping $St$) are not shown due to limited space. The DNS data from Ireland et al. (2016a) are plotted at $St$ matching the experiment, but solely at $R_\lambda = 398$. As stated earlier, since $\langle w_r^- \rangle$ from DNS is insensitive to $R_\lambda$ between $R_\lambda = 88 - 598$ in the $St$ and $r$ ranges we are interested in (and specifically DNS results between $R_\lambda = 224$ and 398 are near-identical), DNS data at $R_\lambda = 398$ with $St$ matching experimental conditions are sufficient for comparison.

From figure 4, we noticed that the experimental data of $\langle w_r^- \rangle$ versus $r$ closely match with DNS data within experimental uncertainty (5%~10%) at all experimental conditions, provided that $r \gtrsim 10\ \eta$. In each case, $\langle w_r^- \rangle$ increases with decreasing slope when $r$ increases. When $r \lesssim$ $10\,\eta$, the experimental results are consistently higher than DNS for all experimental conditions. In order to visualize this discrepancy more clearly, we plot the same experimental conditions in figure 4 using logarithmic scales in figure 5. From figure 5, we see that the value of $\langle w_r^-\rangle/u_\eta$ in the experiment is around 0.37 (with $\pm 0.12$ fluctuations) higher than DNS values when $r = \eta$ for all experimental conditions. This difference does not have a preferred trend when $R_\lambda$ or $St$ are swept in Experiment A and B, respectively.

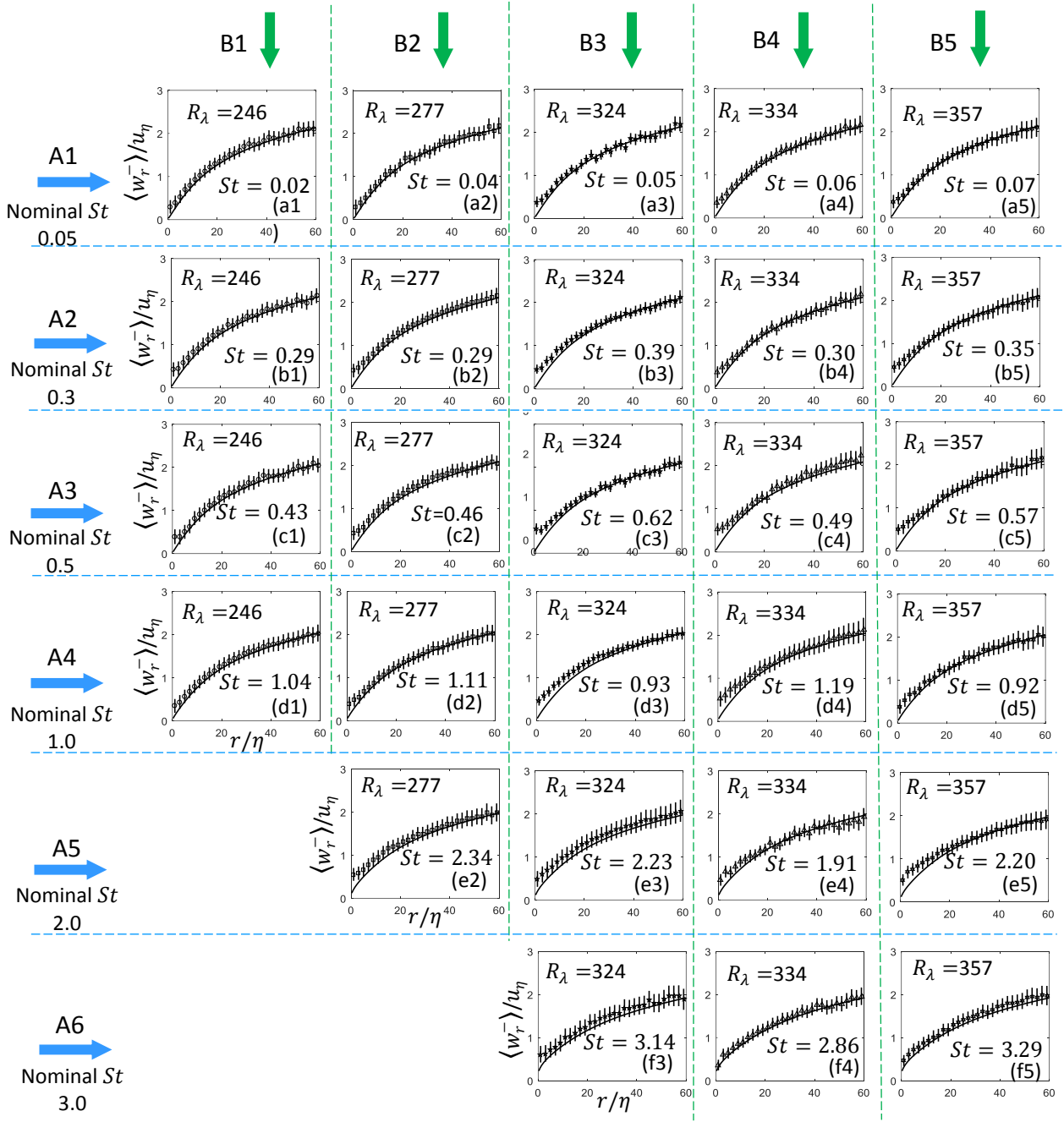

**FIGURE 4**. Mean inward particle-pair RV $\langle w_r^- \rangle$ (normalized by Kolmogorov velocity scale $u_\eta$) versus particle-pair separation distance $r$ (normalized by Kolmogorov length scale) at different $R_\lambda$ and $St$ from experimental measurements (data points with error bars) and DNS (curves). The rows (A1 – A6) represent the sweeping of $R_\lambda$ in Experiment A, while columns (B1 – B5) represent the sweeping of $St$ in Experiment B. For clarity, only 27 experimental conditions between Experiments A and B are plotted, and we omit every other experimental data point in each plot. DNS was taken from Ireland et al. (2016a) under $R_\lambda = 398$ and the matching nominal $St$ values.

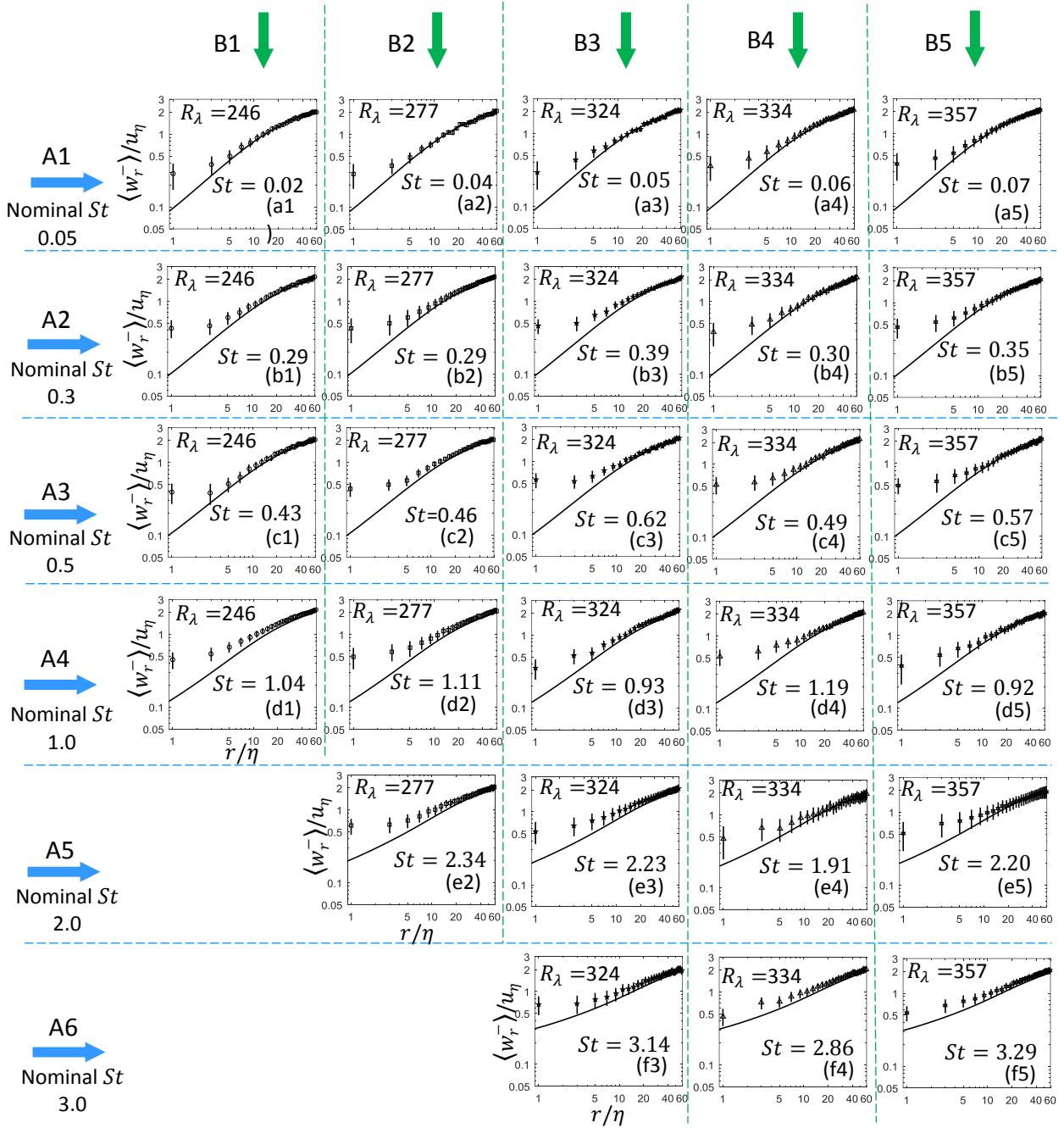

FIGURE 5. Replot of figure 4 using log scale: Mean inward particle-pair RV $\langle w_r^- \rangle/u_\eta$ versus particle-pair separation distance $r/\eta$ at different $R_\lambda$ and $St$ from experiment and DNS.

The persistent discrepancy between experiment and DNS results of $\langle w_r^- \rangle$ at small $r$ through all test conditions recapitulates the observation by Dou et al. (2017) at two test conditions ($St = 0.09$ and $3.46$, both at $R_\lambda = 357$). As explained in Dou et al. (2017), particle polydispersity and finite light sheet thickness effects were the two main contributors to this discrepancy, and contributions from these two effects were comparative. In addition, we noticed that the magnitudes of the discrepancies were very similar among all experimental conditions. This further indicates that the polydispersity and light sheet thickness effects may be the main contributors to the discrepancy as this two effects could be approximate uniform among different test conditions, given the factor that particle size distributions in different size ranges and light sheet thicknesses in different experimental runs were similar to each other.

*3.2 Reynolds Number Effect on Particle-pair RV*

In Test Groups A1-A6, we are able to examine the behavior of $\langle w_r^- \rangle$ versus $r$ as we sweep $R_\lambda$ under similar $St$ in figure 4. To more clearly visualize any effect of Reynolds number on $\langle w_r^- \rangle$, we superimpose all curves of the same nominal $St$ but different $R_\lambda$ together, and show the results in figure 6 and 7, where each plot is a combination of all plots in one row in figure 4 and 5, respectively. The DNS data at $St$ matching nominal $St$ values and $R_\lambda = 398$ are also plotted along with the experimental data. In general, the experimental data points do not show any systematic variation with Reynolds number. Some of the results in figure 7 do seem to reveal a weak trend with $R_\lambda$ at small $r$, e.g. Groups A5 and A6 in figure 7 seem to reveal a decrease in $\langle w_r^- \rangle$ as $R_\lambda$ is increased for fixed $St$. However, such apparent trends should be interpreted with caution since the variation with $R_\lambda$ falls within the experimental uncertainty.

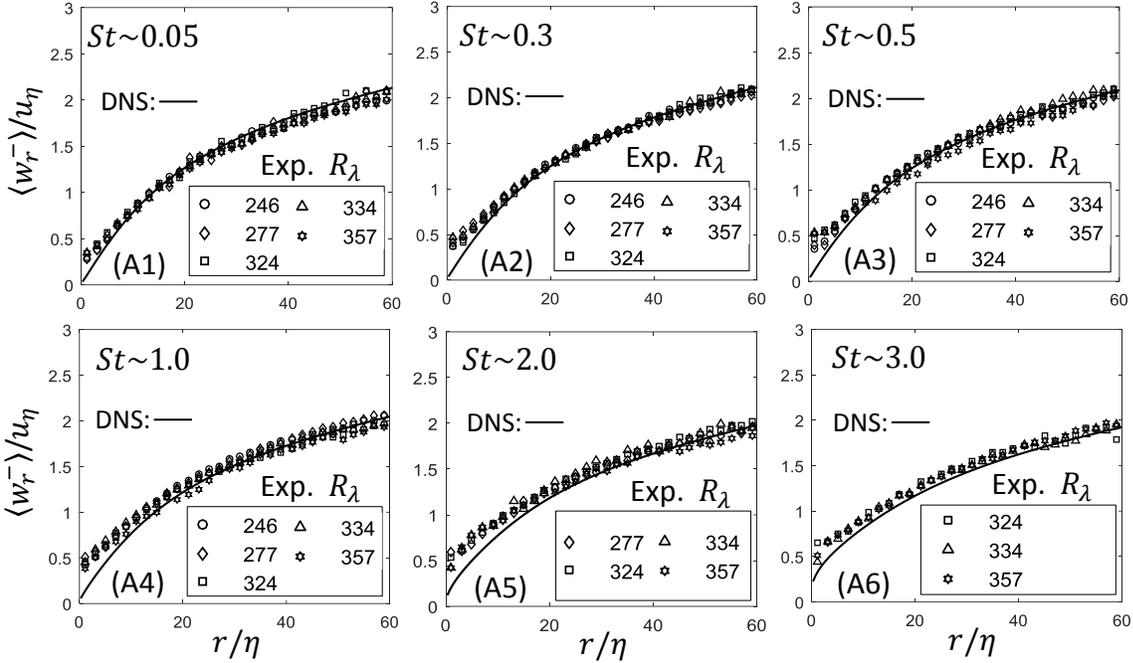

**FIGURE 6**, Superimposition of experimental results of all $R_\lambda$ under the same nominal $St$ value for Test Groups A1 – A6. The DNS data are plotted as a solid line at a $St$ matching the nominal $St$ of the test group with $R_\lambda = 398$. For clarity, we omit every other experimental data point in each plot.

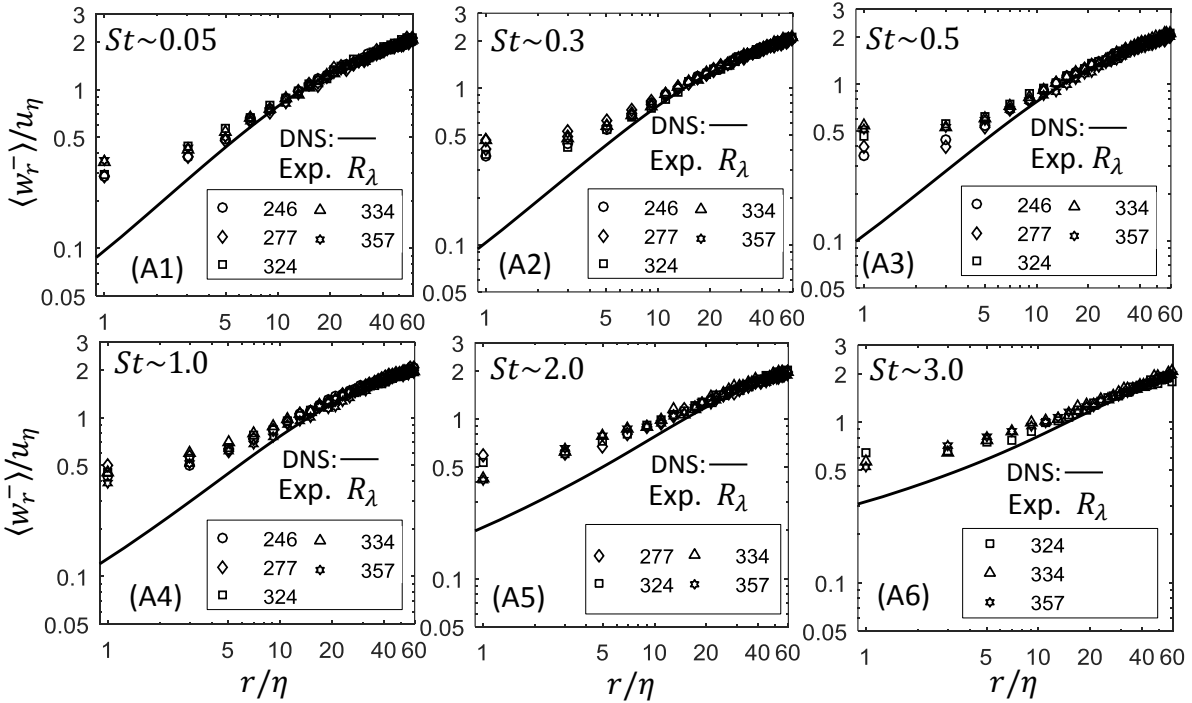

**FIGURE 7**, Replot of figure 6 using log scale: Superimposition of experimental results of all $R_\lambda$ under the same nominal $St$ value for Test Groups A1 – A6.

Furthermore, we plot $\langle w_r^-\rangle/u_\eta$ versus $R_\lambda$ directly at four different particle separation distances ($r = 1\eta, 10\eta, 30\eta,$ and $60\eta$) in figure 8 for all six different $St$. It is seen that $\langle w_r^-\rangle/u_\eta$ is nearly independent of $R_\lambda$, showing no systematic variation of $\langle w_r^-\rangle$ with increasing $R_\lambda$.

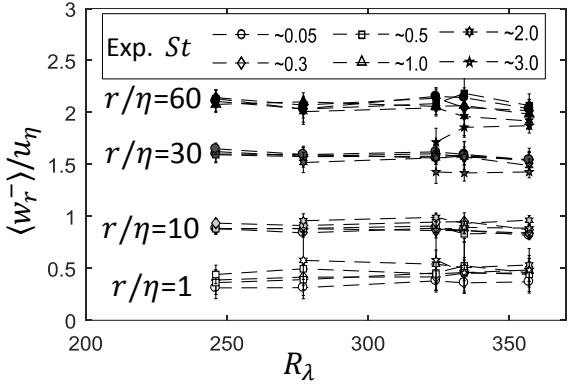

FIGURE 8. Normalized mean inward particle-pair RV $\langle w_r^-\rangle/u_\eta$ against $R_\lambda$ at four different particle separation distances. $r = 1\eta, 10\eta, 30\eta,$ and $60\eta$. Here, $\langle w_r^-\rangle/u_\eta$ has no preferred change when $R_\lambda$ increases, regardless of $St$.

The weak Reynolds number dependence in the range of experimental observations supports previous conclusions from numerical simulations (Bec et al., 2010; Ireland et al., 2016a). Explanations for this behavior were given in Ireland et al. (2016a), which we now summarize. For $St \leq O(1)$, the memory timescale of the inertial particles is sufficiently small such that when $r$ lies in the dissipation range, the inertial particles are only weakly affected by their memory of their interaction with the inertial range turbulence. As such, $St \leq O(1)$ particles are only weakly affected by the trivial Reynolds number effect when $r$ lies in the dissipation range, whereas $St > O(1)$ particles will be affected since they retain a memory of the inertial range scales, whose properties (such as its extent) strongly depend upon $R_\lambda$.

Outside of the dissipation range, for sufficiently large $St$, the filtering mechanism plays an important role in determining the RV behavior. For a given $r$, the effect of filtering depends upon the local Stokes number $St$, and this is directly affected by the trivial $R_\lambda$ effect (i.e. scale separation). In particular, for fixed $r$, $St$ (and hence the filtering effect) decreases as $R_\lambda$ is increased (fixed $\nu$), as observed in Ireland et al. (2016a). This variation is not apparent in our experimental results, which is most likely because outside of the dissipation range, the response times of the particles $\tau_p$ in our experiment are too small compared with the local eddy turn over time $\tau_r$ for the filtering mechanism to be effective (Ireland et al. (2016a) considered particles with $St \leq 30$, whereas we only have $St \leq 3$ in our experiment).

Concerning the non-trivial $R_\lambda$ effect, since the low-order moments of the fluid velocity difference field are (when normalized by the Kolmogorov scales) essentially independent of $R_\lambda$ in the dissipation range (implying they are weakly affected by intermittency), then the low-order inertial particle RV statistics are also essentially independent of $R_\lambda$. For the higher-order RV statistics, not investigated in this paper, this is probably not the case because of the increased intermittency of the turbulent velocity field as $R_\lambda$ is increased.

Ireland et al. (2016a) did detect a weak $R_\lambda$ trend in the dissipation range low-order particle-pair RV using DNS with monodispersed particles. For example, they found a slight decrease of the mean inward particle-pair RV with $R_\lambda$ when $St \approx O(1)$. They attributed this to the fact that the effect of the path-history mechanism depends not only upon the value of $St$, but also the size of the Lagrangian correlation timescales of the fluid velocity gradient measured along the inertial particle trajectories. Their DNS results showed that these timescales decreased slightly with increasing $R_\lambda$ when $St \approx O(1)$ (a non-trivial $R_\lambda$ effect resulting from changes to the spatio-temporal structure of the turbulence when $R_\lambda$ is changed) which decreases the effect of the path-history mechanism, leading to the observed reduction in the RV in this regime.

The experimental uncertainty of our results in this regime (5%~10%) is larger than the change of $\langle w_r^-\rangle/u_\eta$ observed in the DNS, and as such we are not able to conclude whether or not our experiments corroborate the weak $R_\lambda$ dependence observed in the DNS when $St \approx O(1)$. We believe studies on further reducing the experimental uncertainty and increasing the acquirable $R_\lambda$ range would be very helpful to support the weak $R_\lambda$ dependence observed in the DNS. Of course, other factors are the particle polydispersity and laser thickness effects in our experiment, which are absent in the DNS results of Ireland et al. (2016a). How these affect the $R_\lambda$ dependence of the inertial particle RV statistics is currently unknown, which is an important question that we intend to address in future work.

*3.3 Stokes Number Effect on Particle-pair RV*

In Test Groups B1-B5, we are able to examine the trend of curve of $\langle w_r^-\rangle$ versus $r$ as we sweep $St$ vertically under the same $R_\lambda$ in figure 4 and 5. In order to compare the curves at different $St$, we superimpose all curves in each test group of the same $R_\lambda$ from Experiment B and plot the vertically combined plots in figure 9 and 10, respectively. This allows us to further visualize the overall change of the curve slopes. On the combined $\langle w_r^-\rangle/u_\eta$ versus $r/\eta$ plot for each test group, we also plot two curves from DNS of $St$ near the upper and lower $St$ bounds of experimental conditions for that group. Comparing the two simulation curves, and comparing among the experimental curves, we notice that the curves of $\langle w_r^-\rangle/u_\eta$ versus $r/\eta$ are changing when $St$ changes, but no obvious relationship can be obtained due the discernibility.

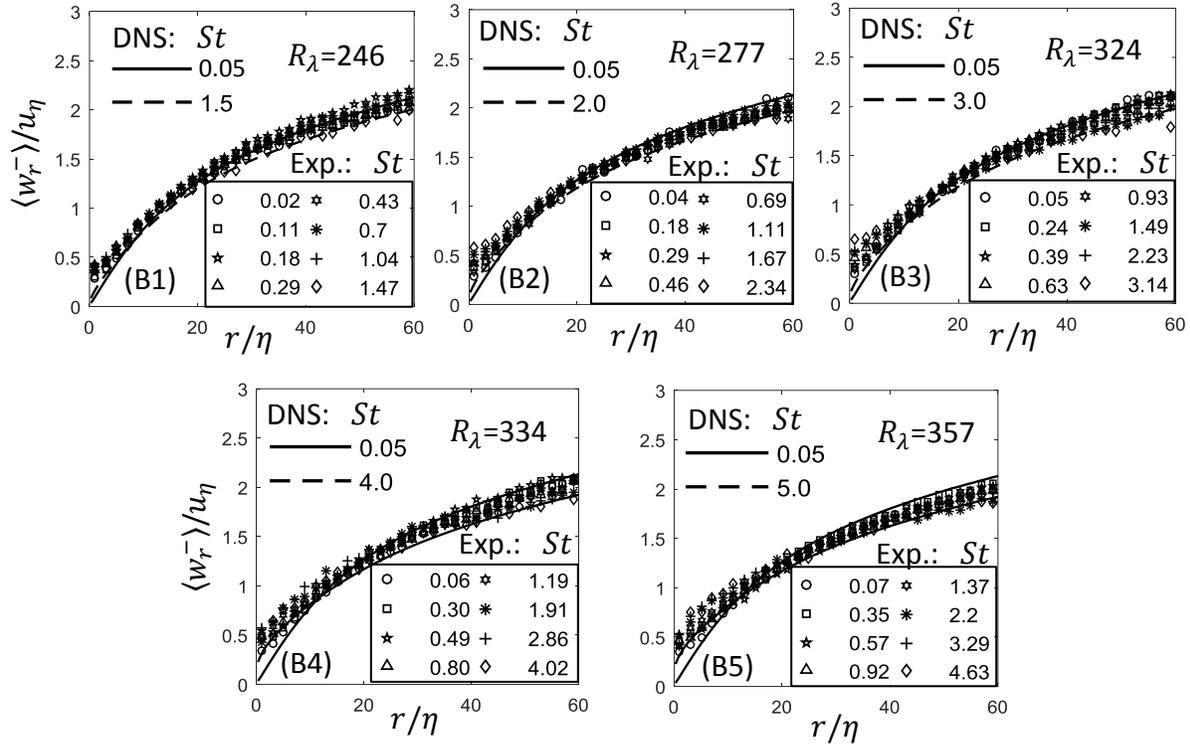

FIGURE 9. Superimposed experimental results of all $St$ under the same $R_\lambda$ for Test Groups B1 – B5. Two DNS curves are plotted as solid and dashed lines at conditions approximate to the highest and lowest $St$ in each test group with $R_\lambda = 398$. For clarity, we omit every other experimental data point in each plot.

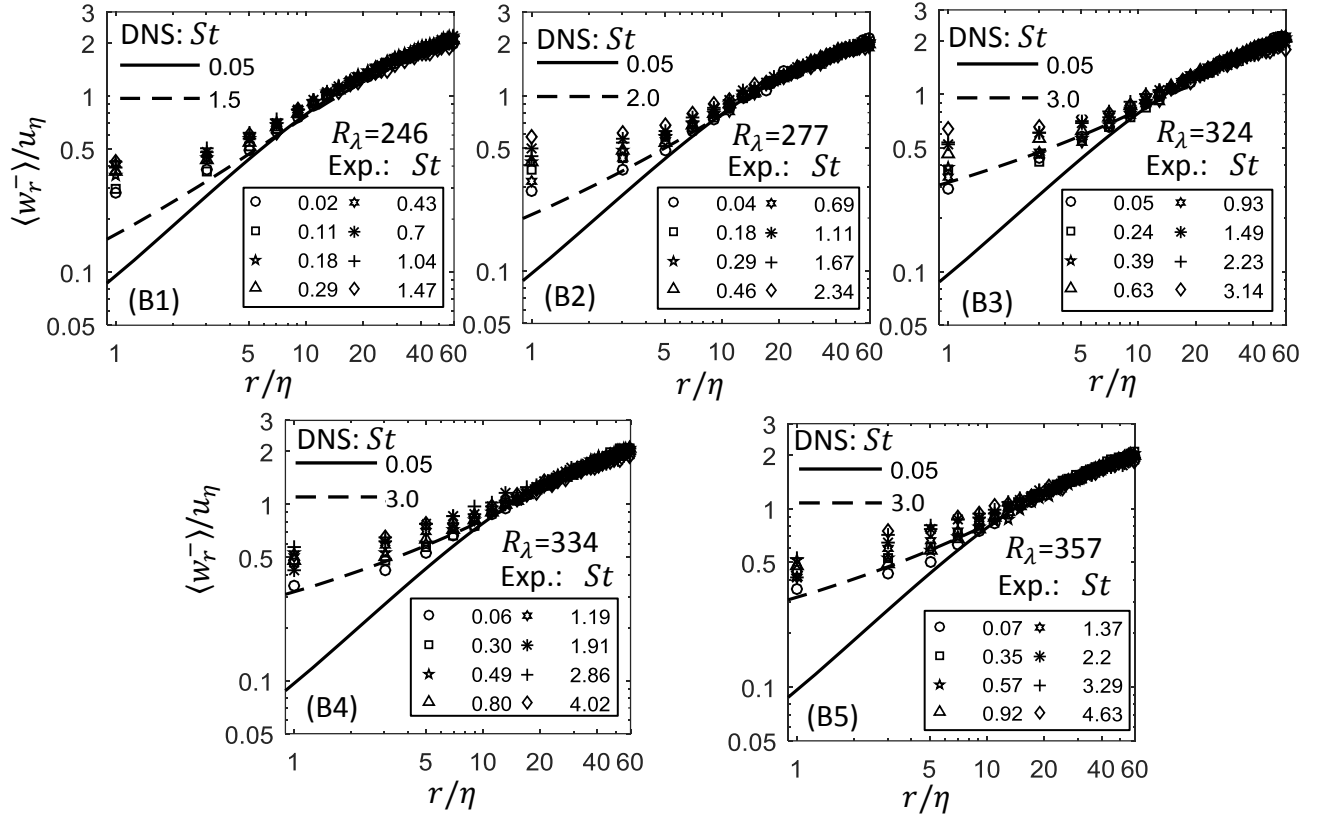

**FIGURE 10.** Replot of figure 9 using log scale: Superimposition of experimental results of all $St$ under the same $R_\lambda$ value for Test Groups B1 – B5.

To clearly visualize this trend, we plot $\langle w_r^- \rangle/u_\eta$ directly against $St$ at single values of $r/\eta$ in figure 11, using experimental and DNS results. Four representative particle separation distances are shown: $r = 1, 10, 30,$ and $60\ \eta$. The experimental values are higher than DNS at $r = 1\ \eta$. When $r$ increases, experimental and DNS results trend to agree with each other to a greater degree, and at $r = 30$ and $60\ \eta$, an excellent match between experiment and DNS are achieved until $St \approx 2$. The improved agreement as $r$ is increased is likey due to the fact that the influence of the particle polydispersity and finite light sheet thickness effects becomes less important as $r$ is increased. Furthermore, both experimental and DNS results indicate that $\langle w_r^- \rangle$ depends more weakly upon $St$ as $r$ is increased. This is again simply a consequence of the fact that for fixed $St$ values, $\tau_r$ become larger than $\tau_p$ as $r$ is increased so that the effect of inertia becomes less and less important.

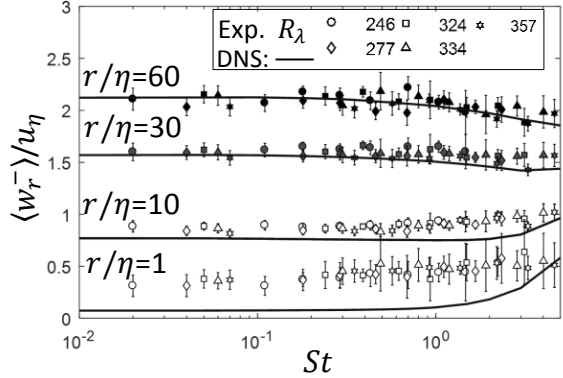

FIGURE 11. Normalized mean inward particle-pair RV $\langle w_r^- \rangle/u_\eta$ against $St$ at four different particle separation distances $= 1\eta, 10\eta, 30\eta,$ and $60\eta$. Black lines represent the DNS results at each corresponding $r/\eta$ and $R_\lambda = 398$. At small $r/\eta$, $\langle w_r^- \rangle/u_\eta$ increases when $St$ increases. At large $r/\eta$, $\langle w_r^- \rangle/u_\eta$ slightly decreases when $St$ increases. $R_\lambda$ do not affect the trend of the data points.

The dependence of $\langle w_r^- \rangle$ on $St$ measured from experiment is consistent with previous numerical studies (Bragg and Collins, 2014b; Salazar and Collins, 2012b) as well as the precursor of the current experimental study by Dou et al. (2017). At small $r$, $\langle w_r^- \rangle$ increases with particle inertia. This is because in this regime the path-history mechanism dominates. At large $r$, particle-pair RV decreases with increasing $St$. This decrease is because as $r$ increases, the filtering mechanism begins to dominate over the path-history mechanism, and the filtering mechanism always reduces the RV.

*3.4 Comparison between Experiment, DNS, and Theoretical Model.*

As discussed by Bragg and Collins (2014b), the analytical model of particle-pair RV from Pan and Padoan (2010) provides a more accurate prediction of the DNS results than previous theoretical models. In order to calculate the mean inward particle-pair RV, $\langle w_r^- \rangle$, Pan and Padoan (2010) adopted the assumption that the probability distribution function (PDF) of particle-pair RV is Gaussian, which leads to $\langle w_r^- \rangle = \sqrt{\langle w_r^2 \rangle/2\pi}$. We would like to examine this theoretical model using experimental and DNS results. This will be done in two steps: (1). The comparison of theoretical predicted $\langle w_r^2 \rangle$ and $\langle w_r^- \rangle$ with experimental and DNS results; (2) The validation of the Gaussian assumption of particle-pair RV, which underlines the relation $\langle w_r^- \rangle = \sqrt{\langle w_r^2 \rangle/2\pi}$ using experimental data.

Following Pan's theory, we first calculated $\langle w_r^2 \rangle$ by solving Eq. (3.5) from Pan and Padoan (2010) numerically. We then calculated $\langle w_r^- \rangle$ using the relation $\langle w_r^- \rangle = \sqrt{\langle w_r^2 \rangle/2\pi}$. These quantities were obtained at $St = [0.01 - 5.0]$, $R_\lambda = 398$, and four particles separation values ($r = 1, 10, 30,$ and $60\eta$). We also obtained $\langle w_r^2 \rangle$ versus $St$ and $\langle w_r^- \rangle$ versus $St$ at similar conditions using experimental and DNS data. Results from the theoretical, numerical, and experimental calculations are normalized by the Kolmogorov velocity scale and shown in figure 12.

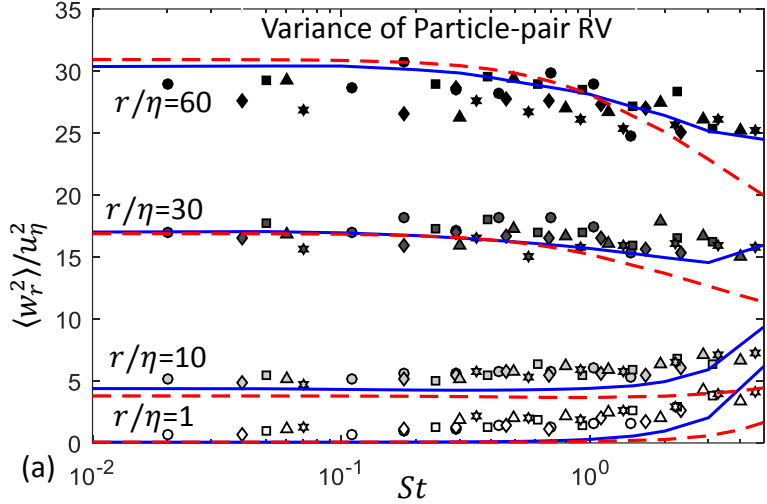

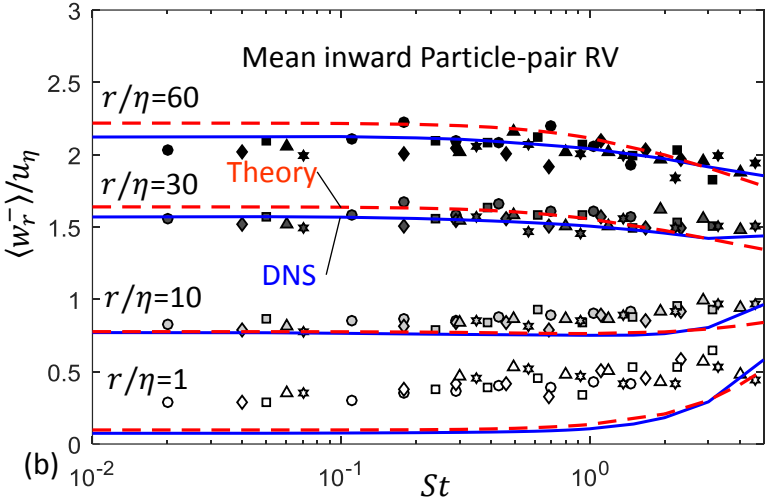

FIGURE 12. Comparison between experimental data, simulation results, and theoretical prediction of normalized variance and mean inward of particle-pair RV versus $St$. Here red lines are the theoretical model predictions, blue curves are the DNS simulation results, and discrete dots are the experimental data. Results are plotted at particle separation distances of $r = 1, 10, 30,$ and $60\,\eta$. (a) Normalized variance of particle-pair RV; (b) normalized mean inward particle-pair RV.

From figure 12, we observe that, in the range $0 < St \lesssim 1.0$, the theoretical prediction for $\langle w_r^2 \rangle$ matches the DNS data well across the range of $r$. The slight discrepancies observed are a consequence of slight errors in the empirical formula used as input to the theory for the fluid second-order structure function. Both the theoretical model and DNS results are smaller than the experimental results at $r \lesssim 10\eta$, but match with experiment result at larger $r$. This is again a result of particle polydispersity and finite light sheet thickness and in the experiment, whose effects are only observable for $r \lesssim 10\eta$. However, when $St \gtrsim 1.0$, the theoretical prediction for $\langle w_r^2 \rangle$ departs from both the experimental and DNS results for all $r$. Indeed, the theoretical results are $20\% - 100\%$ smaller than the DNS and experimental results.

As discussed in Bragg and Collins (2014b), a possible explanation of this disagreement is that in Pan's theoretical model, the particle backward-in-time dispersion is approximated using forward-in-time counterpart. The need to specify the backward-in-time dispersion arises in their model through their approximation for the fluid relative velocities experienced by the inertial particles along their path-history. In this way, the quality of the closure they prescribe for the backward-in-time dispersion will determine, in a significant way, the degree to which their model can describe the path-history mechanism affecting the RV statistics. It was suggested in Bragg and Collins (2014b) that since inertial particles may disperse backward-in-time faster than forward-in-time, the approximation of their equivalence in the Pan theory may explain its under-prediction of the particle-pair RV. In a recent study (Bragg et al., 2016b) it was in fact demonstrated, both theoretically and numerically, that inertial particles separate faster backward-in-time than forward-in-time, and that they may differ by orders of magnitude in the dissipation regime when $St \geq O(1)$. This then supports the explanation given by Bragg and Collins (2014b) for the cause of the under-predictions by the Pan theory.

Despite the discrepancy observed in figure 12 for $\langle w_r^2 \rangle$, the theoretical predictions for $\langle w_r^- \rangle$ compare more favorably with the DNS results. A possible explanation for this is related to the Gaussian assumption used in the theory to obtain a prediction for $\langle w_r^- \rangle$ from $\langle w_r^2 \rangle$, namely $\langle w_r^- \rangle = \sqrt{\langle w_r^2 \rangle / 2\pi}$. This relationship is exact when the PDF of $w_r$ is Gaussian, however, we expect that this PDF will be non-Gaussian (Ireland et al., 2016a). To test the validity of $\langle w_r^- \rangle = \sqrt{\langle w_r^2 \rangle / 2\pi}$ we plot the ratio $\langle w_r^- \rangle / \sqrt{\langle w_r^2 \rangle / 2\pi}$ as a function of $r$ from Test Groups A4 and B5 in figure 13 (a) and (b), respectively. It is evident that the value of $\langle w_r^- \rangle / \sqrt{\langle w_r^2 \rangle / 2\pi}$ is always less than unity based on our experimental data, and it monotonically decreases with decreasing $r$ from 0.97 to 0.80. Our experiment is not be able to obtain reliable data for $r < \eta$, but the DNS results reported by Ireland et al. (2016a) reveal that this ratio will be even lower when $r < \eta$ ($\langle w_r^- \rangle / \sqrt{\langle w_r^2 \rangle / 2\pi} \approx 0.4$ when $r = 0.25\eta$). The conclusion then is that the Gaussian result $\langle w_r^- \rangle = \sqrt{\langle w_r^2 \rangle / 2\pi}$ would lead to an over-prediction of $\langle w_r^- \rangle$ when the correct value for $\langle w_r^2 \rangle$ is prescribed, especially at small $r$.

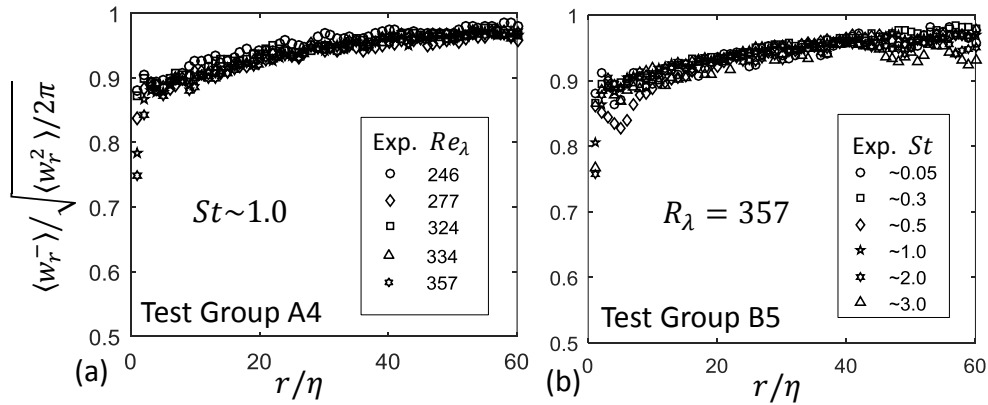

FIGURE 13. The ratio of $\langle w_r^- \rangle$ and $\sqrt{\langle w_r^2 \rangle / 2\pi}$ calculated from experimental data. (a) Ratio calculated from Test Group A4, and (b) ratio calculated from Test Group B5. Note that regardless of the $R_\lambda$ or $St$, the result varies between 0.8-0.97 consistently.

Therefore, although Pan's theory under-predicts the variance $\langle w_r^2 \rangle$ at $St \gtrsim 1$, this under-prediction is compensated for through its use of the use of the Gaussian assumption to compute $\langle w_r^- \rangle$ from $\sqrt{\langle w_r^2 \rangle / 2\pi}$, leading to a much better prediction for $\langle w_r^- \rangle$ than $\langle w_r^2 \rangle$ when $St \gtrsim 1$.

In addition, by improving the spatial resolution (e.g. a higher magnification camera lens), the experimental method used in this study offers a potential of exploring the influence of $R_\lambda$ on particle laden flows at high $R_\lambda$ values and large particles separations which cannot be reached by DNS in the near future.

## 4. Conclusion

We systematically studied, for the first time, the effect of Reynolds number and particle inertia on particle-pair RV through experiments in homogeneous, isotropic turbulence. In Experiments A and B, $R_\lambda$ and $St$ were independently varied, respectively, and we obtained an excellent match of mean inward particle-pair RV $\langle w_r^- \rangle$ between experiment and DNS results throughout the range of experiment conditions, except for particle-pair separations in the dissipation range. The discrepancies in the dissipation range are likely due to particle polydispersity and finite light sheet thickness effects, which were absent in the DNS. We found that Reynolds number has essentially no effect on $\langle w_r^- \rangle$ except within a small region of the parameter space. In this small region, the experimental results show a weak dependence of $\langle w_r^- \rangle$ on the Reynolds number; however, the variation fell within the range of experimental uncertainty, and so must be viewed with caution. We observed that particle inertia enhances particle-pair RV at small particle separation distances, due to the "path-history" mechanism, but particle inertia decreases particle-pair RV at large particle separation distances due to the "inertial filtering" mechanism. The findings are all qualitatively consistent with previous theoretical and numerical results. Lastly, through the comparison between experiment, DNS, and theory, we found that the variance of the particle-pair RV, $\langle w_r^2 \rangle$, predicted by Pan's theoretical model is smaller than the experimental or DNS results when $St \gtrsim 1$. In contrast, the model predictions for $\langle w_r^- \rangle$ compare more favorably with experimental and DNS results when $St \gtrsim 1$. This occurs because the model uses a Gaussian approximation to relate $\langle w_r^2 \rangle$ to $\langle w_r^- \rangle$, which over-estimates $\langle w_r^- \rangle$ (when the correct value of $\langle w_r^2 \rangle$ is prescribed), but since the model under-predicts $\langle w_r^2 \rangle$, the two errors cancel each other out, leading to good predictions for $\langle w_r^- \rangle$.

## Acknowledgement

This work was supported by the National Science Foundation through Collaborative Research Grants CBET-0967407 (HM) and CBET-0967349 (LRC). We thank Dr. Peter Ireland for providing DNS data for comparison with our experiment results. We thank support provided by the Center for Computational Research at the University at Buffalo for Monte Carlo Analysis.